**Quantum memristors for neuromorphic quantum machine learning**

Lucas Lamata

Departamento de Física Atómica, Molecular, y Nuclear, Facultad de Física, Universidad de Sevilla, Apartado 1065, E-41080 Sevilla, Spain

Email Address: llamata@us.es

Quantum machine learning may permit to realize more efficient machine learning calculations with near-term quantum devices. Among the diverse quantum machine learning paradigms which are currently being considered, quantum memristors are promising as a way of combining, in the same quantum hardware, a unitary evolution with the nonlinearity provided by the measurement and feedforward. Thus, an efficient way of deploying neuromorphic quantum computing for quantum machine learning may be enabled.

**Introduction**

Quantum machine learning is an emerging field inside quantum technologies, which may enable to carry out more efficient machine learning calculations with near-term quantum devices [1-4]. It is apparent that a purely quantum unitary evolution is not enough for an efficient learning process, which implies irreversibility to some extent [1,2]. Therefore, most quantum learning algorithms imply some variant of the so-called quantum-classical protocols, such as variational quantum eigensolvers [5-7], quantum Boltzmann machines [8], quantum approximate optimization algorithms [9], and quantum reinforcement learning protocols [10-14]. Besides the basic digital quantum circuit model, all these algorithms include quantum measurements and feedforward in some part of the execution, with subsequent further unitary evolution, further classical part, and so on. Usually, one considers a universal quantum computer based on the digital paradigm for carrying out all these quantum algorithms. However, it may be sensible to employ specialized quantum hardware, which incorporates ab initio both the unitary evolution and the quantum measurement part in the same building block. This is what a quantum memristor [15] may enable.

A quantum memristor [16-20] is a quantum device that is composed of a two-dimensional quantum system, in the most basic instances, together with a weak measurement and feedforward, in the same basic unit. Its properties are that it can sustain a unitary evolution, and in the classical limit it behaves with a hysteresis dynamics, such as a classical memristor [21]. Several of these units may be coupled together to form a network of quantum memristors [22,23], which may be regarded as a neuromorphic quantum architecture. Even though it is not obvious at first sight that this quantum paradigm may carry out more efficient quantum computing protocols than other architectures, it has the basic ingredients of unitary evolution, nonlinearity provided by the measurement, and a scalable architecture. Therefore, quantum memristors could be useful bricks for quantum machine learning devices.

**Quantum memristors for neuromorphic quantum machine learning**

Machine learning aims at having a computing device that can acquire information from the outer world, so called environment, and use that information to improve its performance for some task, in each iteration [24]. This is referred to as learning from the environment. The idea is that interacting with the external world in successive steps, may enable the machine to improve its behaviour for the desired task, sometimes even at superhuman level [25]. Even though classical machine learning already works extremely well, which is evidenced in the multiple daily life applications we enjoy, it is true that processing an increasing amount of data with algorithms that do not scale optimally for this information size is posing a difficulty both in terms of resources and energy consumption. Quantum computers may be a way forward to scale up machine learning computations due to the quantum parallelism, quantum speedup at least for certain kinds of problems, and more efficient energy use in the computing process. However, it is not obvious at first sight which, if any, of the multiple proposed quantum machine learning algorithms may be useful in the short or mid-term for carrying out tasks in better or at least complementary ways to existing classical machine learning protocols.

Some of the quantum machine learning protocols rely on the good connection of quantum systems to linear algebra, via, e.g., the quantum phase estimation algorithm, efficient distance computations, etc [2]. This kind of algorithms normally rely on a single quantum evolution followed by a final measurement. However, they often require a very large number of qubits which is not accesible today or in the near future. Other kinds of quantum machine learning protocols, which have less rigorous evidence of speedup, but nevertheless may be carried out with noisy intermediate-scale quantum computers, are included in the paradigm of classical-quantum protocols. This consists of an iteration of several instances of quantum evolution followed by a measurement, where the subsequent quantum evolution depends on the outcome of the previous measurement, via feedforward. This includes variational quantum eigensolvers, quantum Boltzmann machines, quantum approximate optimization algorithms, quantum reinforcement learning, and quantum memristors.

Of all the existing quantum paradigms for classical-quantum protocols, an appealing of quantum memristors is its in-built modular structure combining in the same unit the quantum coherence via the two-level quantum system and the measurement part, necessary for the nonlinear evolution and the learning process. This seems convenient for scaling up these kinds of protocols involving several iterations of quantum and classical dynamics with intermediate measurements. This is at variance, for example, with the field of quantum simulations, where normally one carries out a single quantum dynamics followed by a final measurement, and more standard quantum architectures already perform very efficiently [26].

It is hard to really predict whether this quantum paradigm of a network of quantum memristors for neuromorphic quantum machine learning will perform better than classical machine learning protocols such as standard neural networks, but this is also a usual problem when comparing a neural network with another one in the classical realm. We, as humans evolved from animals, deal with a very reduced amount of degrees of freedom in our conscious thoughts, as compared with the amount of neurons we have in our brains. Nevertheless, we are perfectly aware of the capabilities of our brains to carry out conscious and useful thinking and knowledge acquisition, which enables us to live in the technological world we have collectively created. I cannot really predict if we will ever be able to understand the inner workings of our brains [27] or of an increasingly complex neural network technology, either classical or quantum. Perhaps it may not even be possible physically or from a computability point of view, but it may not even matter. From a purely pragmatic point of view, if we can use increasingly complex neural networks for an increasingly more advanced set of uses, we may enjoy of unprecedented capabilities for enhancing both our species and the other living species well-being.

**Possible implementations with current quantum devices**

Implementations of quantum memristors with current quantum platforms have been proposed in quantum photonics, superconducting circuits, trapped ions, and polaritons [20]. An experimental implementation in a quantum photonics chip has evidenced that the concept of quantum memristor can be realized in real quantum systems, behaving as expected [19]. Other platforms such as superconducting circuits and trapped ions are also promising for this purpose, and the emergence, in the past few years, of Rydberg atoms as a scalable and controllable quantum platform may also suggest that this concrete platform could be successfully employed for this purpose [28]. The same protocols employed in the fast-developing realm of scalable quantum error correction, for error detection and correction via measurement and feedforward [29], may be directly employed for deploying a scalable network of quantum memristors. Thus, the paradigm of scalable neuromorphic quantum machine learning seems plausible to realize in the mid-term with foreseeable technology [30]. This could also be combined with the tools of digital-analog quantum simulations, computing, and machine learning, which may enable faster scalability and quantum control combined [31,32].

**Conclusions**

Quantum machine learning may enable quantum devices to carry out more efficient machine learning tasks in the near term. To this aim, the combination of unitary evolution with classical optimization, in sequential iterative steps, named quantum-classical protocols, seems a promising path for noisy-intermediate scale quantum computers. In this way, the nonlinear character of the quantum measurement process provides a crucial ingredient for a quantum learning beyond what purely reversible unitary evolution could provide. Among the diverse quantum-classical paradigms appearing for quantum machine learning, quantum memristors seem promising because of their modular structure that combines in the same hardware unit, the quantum memristor, both the unitary and measurement parts, as well as the feedforward. In this way, and similarly to specialized classical neuromorphic hardware, which is expected to implement neural networks in a more efficient way, quantum memristors may enable one to scale up quantum neural networks and quantum neuromorphic computing more successfully and enable neuromorphic quantum machine learning with near term devices. This may permit to scale up machine learning calculations for those tasks which could be properly mapped to this emerging quantum paradigm. Finally, the concept of a quantum memristor network may be also related to the emerging paradigm of digital-analog quantum computation and simulation, and its connection to machine learning, namely, digital-analog quantum machine learning. This is apparent, as quantum memristors are discrete units, with digital character, and at the same time they have analog ingredients in their execution, such as the weak measurement or continuous time unitary dynamics in each unit. This could allow one for both faster scalability and versatility of the device.

**Acknowledgements**

I acknowledge the support from grants PID2022-136228NB-C21 and PID2022-136228NB-C22 funded by MCIN/AEI/10.13039/50110001103 and "ERDF A way of making Europe". This work has also been financially supported by the Ministry for Digital Transformation and of Civil Service of the Spanish Government through the QUANTUM ENIA project call - Quantum Spain project, and by the European Union through the Recovery, Transformation and Resilience Plan - NextGenerationEU within the framework of the "Digital Spain 2026 Agenda".

**References**

[1] Yunfei Wang and Junyu Liu, A comprehensive review of quantum machine learning: from NISQ to fault tolerance, Rep. Prog. Phys. **87**, 116402 (2024).

[2] J. Biamonte et al., Quantum machine learning, Nature **549**, 074001 (2017).

[3] L. Lamata, Quantum machine learning and quantum biomimetics: A perspective, Mach. Learn.: Sci. Technol. **1**, 033002 (2020).

[4] L. Lamata, Quantum Machine Learning Implementations: Proposals and Experiments, Adv. Quantum Technol. **6**, 2300059 (2023).

[5] A. Peruzzo et al., A variational eigenvalue solver on a photonic quantum processor, Nature Comm. **5**, 4213 (2014).

[6] J. Casanova et al., From transistor to trapped-ion computers for quantum chemistry, Sci. Rep. **4**, 3589 (2014).

[7] J. Tilly et al., The Variational Quantum Eigensolver: A review of methods and best practices, Phys. Rep. **986**, 1 (2022).

[8] M. H. Amin, E. Andriyash, J. Rolfe, B. Kulchytskyy, and R. Melko, Quantum Boltzmann machine, Physical Review X **8**, 021050 (2018).

[9] E. Farhi, J. Goldstone, S. Gutmann, A quantum approximate optimization algorithm, arXiv:1411.4028 (2014).

[10] L. Lamata, Basic protocols in quantum reinforcement learning with superconducting circuits, Sci. Rep. **7**, 1609 (2017).

[11] F. Albarrán-Arriagada, J. C. Retamal, E. Solano, and L. Lamata, Measurement-based adaptation protocol with quantum reinforcement learning, Physical Review A **98**, 042315 (2018).

[12] S. Yu et al., Reconstruction of a photonic qubit state with reinforcement learning, Adv. Quantum Technol. **2**, 1800074 (2019).

[13] M. L. Olivera-Atencio, L. Lamata, M. Morillo, and J. Casado-Pascual, Quantum reinforcement learning in the presence of thermal dissipation, Phys. Rev. E **108**, 014128 (2023).

[14] V. Dunjko, H. J. Briegel, Machine learning & artificial intelligence in the quantum domain: a review of recent progress, Rep. Prog. Phys. **81**, 074001 (2018).

[15] L. Lamata, Memristors go quantum, Nature Phot. **16**, 265 (2022).

[16] P. Pfeiffer, I. L. Egusquiza, M. Di Ventra, M. Sanz, and E. Solano, Quantum Memristors, Sci. Rep. 6, 29507 (2016).

[17] J. Salmilehto, F. Deppe, M. Di Ventra, M. Sanz, and E. Solano, Quantum Memristors with Superconducting Circuits, Sci. Rep. **7**, 42044 (2017).

[18] M. Sanz, L. Lamata, and E. Solano, Quantum Memristors in Quantum Photonics, APL Photonics **3**, 080801 (2018).

[19] M. Spagnolo et al., Experimental photonic quantum memristor, Nature Phot. **16**, 318 (2022).

[20] P. A. Forsh, S. Y. Stremoukhov, A. S. Frolova, K. Y. Khabarova, and N. N. Kolachevsky, Quantum memristors: a new approach to neuromorphic computing, Physics-Uspekhi **67,** 855 (2024).

[21] M. Di Ventra and Y. Pershin, Memristors and Memelements (Springer Nature, 2023).

[22] S. Kumar et al., Entangled quantum memristors, Phys. Rev. A **104**, 062605 (2021).

[23] S. Kumar et al., Tripartite Entanglement in Quantum Memristors, Phys. Rev. Applied **18**, 034004 (2022).

[24] S. Russell and P. Norvig, Artificial Intelligence: A Modern Approach (Pearson, 2020).

[25] D. Hassabis, Nobel Lecture, https://www.nobelprize.org/uploads/2024/12/hassabis-lecture.pdf

[26] I. M. Georgescu, S. Ashhab, and F. Nori, Quantum simulation, Rev. Mod. Phys. **86**, 153 (2014).

[27] https://www.humanbrainproject.eu/en/

[28] A. Browaeys, and Thierry Lahaye, Many-body physics with individually controlled Rydberg atoms, Nature Phys. **16**, 132 (2020).

[29] D. Bluvstein et al., Logical quantum processor based on reconfigurable atom arrays, Nature **626**, 58 (2024).

[30] https://www.pasqal.com, https://www.quera.com/

[31] L. Lamata, A. Parra-Rodriguez, M. Sanz, and E. Solano, Digital-Analog Quantum Simulations with Superconducting Circuits, Adv. Phys.: X **3**, 1457981 (2018).

[32] L. Lamata, Digital-Analog Quantum Machine Learning, arXiv:2411.10744 (2024).